# Shining a light on Spotlight: Leveraging Apple's desktop search utility to recover deleted file metadata on macOS

Tajvinder Singh Atwal [a, b], Mark Scanlon [b, c, *], Nhien-An Le-Khac [b, c]

[a] *Financial Conduct Authority, London, United Kingdom*
[b] *School of Computer Science, University College Dublin, Ireland*
[c] *Forensics and Security Research Group, University College Dublin, Ireland*

ABSTRACT

Spotlight is a proprietary desktop search technology released by Apple in 2004 for its Macintosh operating system Mac OS X 10.4 (Tiger) and remains as a feature in current releases of macOS. Spotlight allows users to search for files or information by querying databases populated with filesystem attributes, metadata, and indexed textual content. Existing forensic research into Spotlight has provided an understanding of the metadata attributes stored within the metadata store database. Current approaches in the literature have also enabled the extraction of metadata records for extant files, but not for deleted files. The objective of this paper is to research the persistence of records for deleted files within Spotlight's metadata store, identify if deleted database pages are recoverable from unallocated space on the volume, and to present a strategy for the processing of discovered records. In this paper, the structure of the metadata store database is outlined, and experimentation reveals that records persist for a period of time within the database but once deleted, are no longer recoverable. The experimentation also demonstrates that deleted pages from the database (containing metadata records) are recoverable from unused space on the filesystem.

© 2019 The Author(s). Published by Elsevier Ltd. This is an open access article under the CC BY-NC-ND license (http://creativecommons.org/licenses/by-nc-nd/4.0/).

## Introduction

The popularity and market share of Apple Macintosh computers has increased over the last decade and is reflected in the number of Apple devices being submitted to digital forensic laboratories. During this same period, the support for macOS (previously Mac OS X) on-disk artefacts has also increased as a result of the greater level of research being conducted and the number of Mac compatible forensic examination tools available. Despite this upward trend, little research has been conducted concerning Apple's system wide desktop search technology named Spotlight.

The requirement for quickly and efficiently locating specific data has become more important as the amount of digital data being stored by individuals and companies has increased (Du et al., 2017). The two most popular operating systems: Microsoft Windows and Apple macOS have historically provided filesystems designed for organising files within a hierarchical directory structure. Retrieval of files is reliant on remembering where the data was saved, what it was named or how it was organised. Users are able to search by using a limited set of data attributes, e.g., file name, path, size, date modified, etc. In the early 2000s, desktop search tools were created coinciding with Internet search engines' growth in popularity. These tools offered owners of desktop computers methods to search the contents of their computers and find files based on keywords, attributes, and metadata, and would include searching each file's textual content.

Spotlight is a proprietary desktop search technology developed by Apple for Mac OS X 10.4 (Tiger) and is still included with current versions of macOS. Spotlight allows users to search for files or information by querying databases populated with filesystem attributes, metadata, and indexed textual content (Apple Inc. and Spotlight O, 2013). Announced at the Apple Worldwide Developers Conference (WWDC) in 2004, it has been integrated with every major release of the operating system since Mac OS X version 10.4 (Tiger).

The contribution of this paper consists of a novel approach to examining Spotlight metadata stores through two new strategies;

\* Corresponding author. School of Computer Science, University College Dublin, Ireland.
*E-mail addresses:* tajvinder.atwal@gmail.com (T.S. Atwal), mark.scanlon@ucd.ie (M. Scanlon), an.lekhac@ucd.ie (N.-A. Le-Khac).

extracting persistent records of deleted files directly from the Spotlight database and recovering records from deleted database pages residing within unused space of the filesystem. This paper also provides proof-of-concept implementation, based on an understanding of the database structure, which support the extraction of records from these sources.

The rest of paper is organised as follows: the background and related work to the research presented in the paper is discussed in Section 2 and in Section 3 respectively. The approach used for this work is highlighted in Section 4. Experimentation and analysis is outlined in Section 5. We discuss the experimental results in Section 6. Finally, we conclude and discuss on future work in Section 7.

**Background**

*Spotlight Metadata Server*

Spotlight's core function is to enable users to find data on the system by providing methods to search and identify files from filesystem attributes, indexed file metadata, and live searches. Out of the box, Spotlight is integrated with the standard suite of Apple applications including Address Book, Calendar, Finder, Mail, Picture files, System Preferences (Apple Inc. and. Spotlight O, 2013). The Spotlight Metadata Server (MDS) is the core Spotlight management service and indexing server, which forms part of the CoreServices framework, running as a background process daemon in macOS. The MDS is managed at the kernel level to run at start-time, although the MDS runs outside the kernel abstraction layer (Apple Inc. and. Daemons and, 2016). The application files for the service are located within the CoreServices Framework directory. The MDS is notified of all changes within a directory (by FSEvents). If a file has been created, modified, or deleted, it will launch the appropriate metadata importer responsible for extracting contextual text and embedded metadata before it is passed back to the MDS for indexing. The Content Index and Metadata Store database (store.db) are then populated with the indexed material. Searches used to query these stores are sent via the MDS from either the Spotlight graphical user interface (GUI), commands run on the terminal, or by applications making use of Apple's Metadata Framework (J. and L.. MacX and i, 2012). Spotlight is designed to be extensible. Application developers are able to integrate their custom software and file formats by making use of the Spotlight search index and developing custom metadata importers via the Metadata Framework API methods in order to query and update the index (Apple Inc. and. Spotlight O, 2013).

*File Systems Events*

File Systems Events (FSEvents) is Apple's filesystem notification API that provides a framework for detecting/recording changes that have occurred within the directory hierarchy. The technology passively monitors changes at the kernel level (Kaushik, 2007) and forms the foundation of Apple's backup technology, Time Machine. Changes monitored include; the modification of a file's filesystem attributes (e.g., metadata such as timestamps), the content change caused by modifying a file within a directory or file/folder deletion. Spotlight is a subscriber to FSEvents notifications, ensuring that the MDS is notified of any changes (Apple Inc. and. File System, 2012). FSEvents runs at the kernel level and notifies the MDS of any changes made to a file's metadata. This will result in a Metadata Importer to be reactivated and extract/populate the database stores with the latest information. The exception to this rule is if the file has been edited on a non-macOS system and then reintroduced. As FSEvents will not have been updated of a change, it will not notify Spotlight. A list of events that are actively monitored by FSEvents is detailed within the FSevents.h source file (Apple Inc. and. Apple Open, 2006) and has also been described by Amit (2006).

*Metadata store database*

The metadata store database (store.db) is used as Spotlight's central database repository for storing extracted metadata values. It is a proprietary database format whose structure is not published or described. The mdimport command populates the store by initialising the MDS to extract metadata key–value pairs (J. and L.. MacX and i, 2012). The metadata store is located within the. Spotlight-v100 folder. The precise location has changed with version changes of macOS.

**Related work**

Metadata is of great interest to digital forensic investigations. However, to make the most if it, the method metadata is stored within files and data repositories must be understood, tested, and shown to be to be reliable. Accessing and understanding device metadata can assist the investigators in examining digital artefacts from filesystems (Du et al., 2018), mobile apps (Faheem et al., 2015), and Internet of Things devices (Sayakkara et al., 2018). Sampson (2013) documented the necessity of understanding metadata and offered strategies to assist with this type of review. A summary is provided below:

- Gather documentation from the publisher of the software used to create the file type.
- Test the software and file container to create familiarity.
- Gather information regarding the file type from third parties, e.g., open source software, integrated tools, etc.
- Test by manually editing and manipulating the metadata using hex editors.

A description of how Spotlight functions and the process of creating integrated tools that communicate with the MDS and FSEvents is well documented by Apple Inc (Apple Inc. and. Spotlight O, 2013). The majority of research conducted on Spotlight has focused on the methods provided by this official developer documentation. There is no published research in the literature that reveals the underlying structure of the metadata store or the forensic opportunities offered by recovering records for deleted files within the database and unused space on the filesystem.

Examining application generated databases is not only an important task for forensic analysis of the data generation software tools, e.g., desktop applications and web browsers (Buchholz and Spafford, 2004), mobile applications (Faheem et al., 2015), and cloud services (Farina et al., 2015), but has been identified for desktop metadata curators, such as desktop file search utilities. Turnbull and Slay (Turnbull and B.BSlay, 2006) emphasised the importance of conducting research into the databases used by Google Desktop Search technologies in order for tools to be designed to extract the data for digital forensic investigations. Parsing data from an unknown data format is inherently dangerous as the incorrect linkage of records or incomplete data parsing can cause any data recovered to be misunderstood and within the scope of the forensic investigation and potentially cause an incorrect conclusion to be drawn. The presence and recovery of deleted records are the first step in providing a product that can be used in a court of law, but many arduous hours of testing must be employed to show it is reliable and fit for purpose.

The Windows Desktop Search service is such an example where understanding the structure has led to forensic opportunities in exploiting the recovery and analysis of recoverable deleted records.

The structure was revealed following on from work performed by Metz (2009a) in describing the data structures of the Extensible Storage Engine format, used by Microsoft in storing data (most notably used in Microsoft Exchange Server). Metz then presented his specification describing the structure of the Windows Search Database file (Metz, 2009b). It was noted that each table located within the database was described with a column identifier, column name, and column type.

In another relevant research, Pavlic and Turnbull (Pavlic and Turnbull (2008) presented research aimed at developing a process model and framework that can be used to extract information from desktop search applications and their associated data structures. Chivers and Hargreaves (2011) completed the process by presenting the recovery from the Windows Search Database. The authors argued that the existing tools used similar programmatic interfaces as those that exist with Spotlight, but this strategy does not offer access to deleted records within the unused space, the database or the filesystem. Instead, they offered a solution by employing carving techniques to recover database records.

With regard to Spotlight, in 2005, Singh created a proof of concept tool fslogger to mimic the way the MDS is notified of file system kernel events by creating a tool that logged FSEvents notifications (Singh, 2005). He introduced a subscriber to FSEvents as a proof of concept tool that can be used to assist software developers to research and test their products. The tool itself was one in a line of subsequent products that provided a better understanding of how the macOS operating system and HFS + filesystem operate by tracking changes on the system.

Following on from the work by Singh (2005), Kaushik (2007) investigated Spotlights integration with the macOS filesystem, in particular, the reliability of notifications updating the content index. The paper focused on the FSEvents notification system and investigated the possibility of missed notifications caused by dropped events. If the Spotlight MDS misses a notification, it will not process the file until a new notification is delivered (caused by some other change at the directory level). In experimentation, an FSEvents subscriber was created to create dropped events (caused by a buffer overflow). This confirmed that the textual content within files would not be indexed. Whenever the system is under increased activity, the possibility of missed events increases which would directly influence the outcomes of experiments conducted within this research project. Either records that should be deleted within the metadata store are not or files that have been newly created/modified will not be indexed.

Scott (2008) outlined the methods an attacker could use to conceal themselves within macOS. From a Spotlight point of view, an attacker is able to subvert the indexing and querying process of the MDS by switching services off or preventing files from being indexing by the use of hidden files and AppleDouble files. These methods are also useful from a researcher's point of view. It provides a method to suppress the Spotlight indexing service for particular files/folders.

Joyce and Adelstein (Joyce et al., 2008) developed the tool MEGA (later named Mac Marshal) that provided access to Spotlight metadata (as well as other macOS artefacts). The tool and website are no longer accessible (http://www.macmarshal.com). The authors offered two approaches for analysing Spotlight metadata records, both leveraging the inbuilt APIs that Apple offer. A mounted volume can be queried by making use of the mdfind terminal command and querying the file by combining it with the mdls terminal command, this led them to develop a custom command-line tool named spquery. In addition, they offered a method to access un-indexed files on a mounted system by use of their tool MEGA that included a Spotlight search tool. The NIJ Criminal Justice Electronic Crime Technology Centre of Excellence published an evaluation of the Mac Marshal tool (of ExcellenceC., 2011) and reported that it provided a "rapid search" using the Spotlight file metadata, however noted that the software required the use and implementation of macOS forensic examination machines.

In conclusion, an investigator is limited to using the in-built Spotlight search functions, either by running terminal commands or using Spotlight to search the contents of a suspect drive. Joyce and Adelstein (Joyce et al., 2008) planned for future versions of MEGA that would enable the user to index un-indexed suspect filesystems with the aim of being able to query all files as required. It is not known if this strategy was ever successful.

Varsalone (J. et al., 2008) provided a similar methodology to query the Spotlight metadata store by making use of the Spotlight GUI and mdls terminal command. A read-only forensic image (acquired/converted to a *.dmg image) is mounted as a read-write structure by shadow mounting the device.

504ENSICS Labs announced a tool, Spotlight Inspector, that parsed the Spotlight metadata store database (store.db) offline, on several examination systems (ensics Releases Digital F, 2013). This was the first record of the database being successfully reverse engineered. However, a note on the 504ENSICS website states that the Spotlight Inspector intellectual property has been acquired by Blackbag Technologies, Inc. As such, 504ENSICS Labs no longer offers direct free downloads of the tool. Today, this tool is no longer available nor was their methodology, results or analysis of the database structure published or made publicly available. It is unknown if the tool was able to recover deleted records or parse the content for these records.

Luttgens and Mandia (Luttgens and P.MMandia, 2014) recorded the presence of Spotlight indices on removable media and network shares. The indices stored information related to files/folders within the local storage system. Their undisclosed tests stated that using the mdfind terminal command on the local system would "not return data from the disconnected source" and that an examination of the Metadata Framework showed that records were removed from the content index immediately after a file was deleted. Their conclusion was that data maintained by Spotlight is only useful within a live response context and that there were no tools available that could reliably parse data stored in the Spotlight index. It is not clear what tests the authors performed or if the searches within the content index were performed via the Spotlight GUI. Nor was it established if the metadata store itself was investigated for records of deleted files.

Each of the related techniques employed in the literature involve using a macOS examination machine, mounting the suspect drive to be examined (as read/write) and extract the records using either closed system tools or querying individual records via the command terminal; a slow and laborious endeavour. There is a gap in the research conducted for Spotlight regarding the persistence of records of deleted files within the metadata store. By understanding the structure of the database, it is expected that the discovery of deleted records will be achievable, in addition, it offers the opportunity of recovering records found within unused space of the filesystem.

**Approach**

The objective of this paper is to study the structure of the Spotlight metadata store database in order to assess if metadata records for files deleted on the filesystem persist. The literature survey has highlighted reasons why this should be investigated further: (i) Records may persist within the metadata store whilst they await removal; (ii) Records may never be deleted due to the dropped file system notifications; (iii) Records that are deleted, may survive as unallocated records within the database; (iv) Deleted database pages may be found within unused space on the filesystem. Hence, in this paper, we aim to answer the following research questions:

- Are metadata records for deleted files persistent within the metadata store?
- Are metadata records recoverable from the unused space on the filesystem when (i) the corresponding file(s) have been deleted, (ii) the Spotlight indexes are reset, and (iii) the operating system is updated?

To answer these questions, in the following subsections, we describe our approach of forensic acquisition and analysis of Spotlight metadata.

*Experiments*

In total, eight (8) experiments have been formulated for this research. An examination protocol for each experiment has been created that details the steps, processes and actions to be taken prior to the creation of a snapshot.

**Experiment 1.** Examination of the persistence of metadata records within the metadata store and unused space on the filesystem. These experiments were conducted on virtualised versions of Mac OS X 10.8, 10.11 and macOS 10.12.

**Experiment 2.** Examination of the persistence of metadata records within the metadata store and unused space on the filesystem. These experiments were conducted on virtualised versions of Mac OS X 10.8, 10.11 and macOS 10.12 operating systems. Mounted volumes were formatted as FAT32, exFAT, and HFS+. FAT32 and exFAT are two Windows-based filesystems that a macOS can access. We are only focusing on the Spotlight artefacts that might be stored on these filesystems. More details on exFAT forensics can be found in (Vandermeer et al., 2018).

**Experiment 3.** Examination of the persistence of metadata records within the metadata store and unused space on mounted volumes that are shared across two operating systems. These experiments were conducted on virtualised versions of macOS 10.12 and Windows 10 operating systems. Mounted volumes were formatted as FAT32, exFAT.

**Experiment 4.** Examination of the persistence of metadata records within the metadata store and unused space on system volumes. In particular, when the Spotlight indices are deleted using the mdutil terminal command. This experiment was conducted on a virtualised version of macOS 10.12.

**Experiment 5.** Examination of the persistence of metadata records within the metadata store and unused space on system volumes. In particular, when the Spotlight indices are deleted via the Spotlight management GUI interface located within System Preferences. These experiments were conducted on virtualised versions of macOS 10.12.

**Experiment 6.** Examination of the persistence of metadata records within the metadata store and unused space on a system volume. In particular, when the Spotlight indices are deleted using the mdutil terminal command and re-populated with ever increasing number of files. Experimentation was conducted on virtualised versions of macOS 10.12.

**Experiment 7.** Examination of the creation of metadata records for the purposes of reverse engineering the metadata store structure. These experiments were conducted on virtualised versions of macOS 10.12 and findings tested on all available test metadata stores.

**Experiment 8.** Examination of the persistence of metadata records within the unused space of ten casework forensic images based on the version of OS X installed.

The following versions of macOS have been examined in our experiments: Mac OS X 10.4 (Tiger), 10.5 (Leopard), 10.6 (Snow Leopard), 10.7 (Lion), 10.8 (Mountain Lion), 10.9 (Mavericks), and macOS 10.12 (Sierra). We also used EnCase v.7.16.00.29, FTK Imager v3.4.3.3, and VMWare Fusion v8.5 in the experiments.

*Data acquisition*

At certain steps before taking a snapshot, it is necessary to ensure that the indexing of the volume by Spotlight has been completed. The mdutil terminal command offers a check on the indexing status of a volume, however, the only (non-programmatic) method is to monitor the progress of indexing via the Spotlight GUI interface or the activity monitor, in particular the activity of the MDS process.

Each individual snapshot from the experiments has been acquired by taking a physical image of the virtual hard disk drive (vmdk) snapshot. Each vmdk was set as read-only, acquired, and verified using FTK Imager. The metadata store database (store.db) files were then exported for further examination and the unused space of the filesystem searched using EnCase.

We have also developed a script to extract metadata records from the virtualised environments for the purposes of validation testing. The script takes the start directory as an argument and recursively traverses the structure, each file found is queried using the mdls terminal command. All reports are exported as text files to a specified directory. This script can be used to conduct searches on the native filesystem or any forensic images mounted on an examination machine. It has been designed to work out of the box, without any special requirements (other than it will only work on Mac OS X/macOS).

**Findings and analysis**

Analysis of the data collected during the experiment stage has revealed the basic structure of the metadata store database. This has enabled scripts to be developed that extract and count records containing metadata, either directly from the metadata store, or from database pages found within unused space on the filesystem. The scripts developed have been used in combination with keyword searches in order to establish if records stored within the database are persistent after a deletion event.

The metadata store database (store.db) has been noted as always being co-located with a version named. store.db. Experiments show that these files are not identical copies of each other. When new file events occur, the. store.db version has been found to be updated first with any new metadata records, with the store. db file being populated shortly afterwards. A record by record comparison on the two versions (using MD5 hash analysis) revealed that the difference between the stores is only limited to what files are being updated on the system (created, modified, and deleted). The existence of these two versions is discussed further in the evaluation and discussion section of this paper.

*Database structure*

Three main types of database pages have been identified within the metadata store: Header Page, Map Page, and Data Page. Each of these is identifiable by the 4-byte signature located at relative offset 0. We developed a python script to automate the splitting of database pages based on these landmarks. The minimum number of pages detected within the databases is eight when no files have been indexed. The file metadata records have all been observed to start from page 8 of the database onwards. In general, data within the pages are stored little-endian (least significant byte first), however big-endian values have been noted as existing within individual records. A description of the pages is detailed within the following sub-section of this paper.

*Header Page*

The first page encountered is identified as the database header. In experiments, this page has consistently been 4096 bytes in length. The key records used for parsing the database header are detailed in Table 1 and the example in Fig. 1.

A number of records have been detected within the page after the store. db path, in particular, a block of data which is repeating in nature and suspected to be Timestamp records. These types of records have fallen out of scope for this research and have not been investigated further at this point.

*Map Page*

The second type of page encountered has been identified as the database map. The first 32 bytes provide information regarding the page. Starting at offset 32, each data page is described by 16 bytes. The first 4 bytes declare the size of each data page with the remaining 12 bytes unknown. The key records used for parsing this data are detailed in Table 2 with an example shown in Fig. 2.

*Data pages tables*

The third type of page structure has been identified as the data page as they have been found to contain tabular data fields and records. The key records used for parsing all data pages are found within the first 20 bytes and a summary in Table 3.

*Metadata records (data page type 9)*

Data found within data page type 9 has been found to consistently contain metadata records and exist in a compressed state (at rest on the filesystem). The compression library used by Apple was discovered by tests performed using Linux and python and has consistently made use of the zlib software library net (Zlib compressed data form, 1996). Fig. 3 shows an example of a compressed data payload within a page. Fig. 4 shows the same page decompressed, revealing the metadata records indexed by Spotlight.

The records located within this page are stored back-to-back, with the size marker of the record found at relative offset 0. For example, in Fig. 5, the size of the record is stored as 8A080000 (little-endian). This equates to the data payload for record 1 is stored within the next 2186 bytes after the record size marker. The size marker for record 2 will then be found within the next 4 bytes after record 1 (i.e., at relative offset 2190). Fig. 5 provides a theoretical example of the storage of metadata records within the database page.

We also developed a python script to automate the splitting of records from this page type in order that they can be examined manually or be used to conduct keyword searches. The script was also modified in order to report the number of records found within the page. The metadata records now have been used to reverse engineer the contents of these records in order to understand the data structures. The protocol for Experiment 7 was specifically created for this task, where a limited set of data was populated with clearly identifiable metadata attributes. The first two records found within the metadata store are a property list file defining the configuration and a record detailing the root directory. Even if the volume contains no files, the metadata store will always be populated with these two records. A detailed analysis of one record can be seen in Fig. 6.

**Table 1**
Header page structure.

| Offset | Bytes | Description |
| --- | --- | --- |
| 00 | 4 | Header string recorded as 8tsd |
| 36 | 4 | Database header page size |
| 324 | variable | Path to store.db on volume |

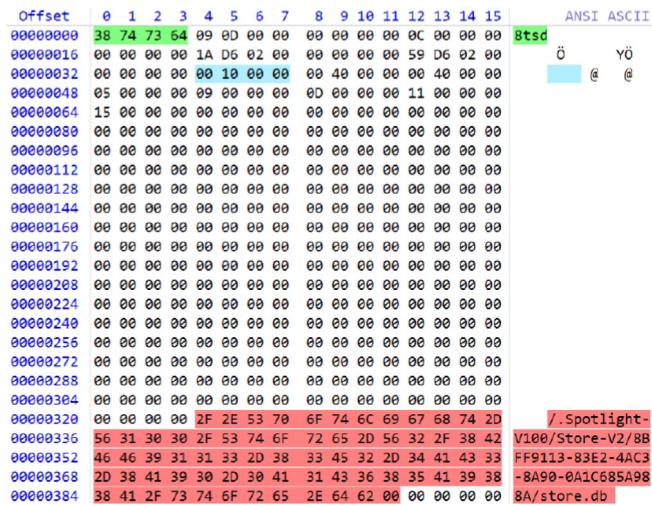

Fig. 1. Example database header.

**Table 2**
Map page structure.

| Offset | Bytes | Description |
| --- | --- | --- |
| 00 | 4 | Header string recorded as 2mbd |
| 04 | 4 | Page size, always 16,384 bytes in length |
| 08 | 4 | Number of pages contained within the map |
| 12 | 4 | Page type record as 12 |
| 16 | 16 | Start of the page map entries |

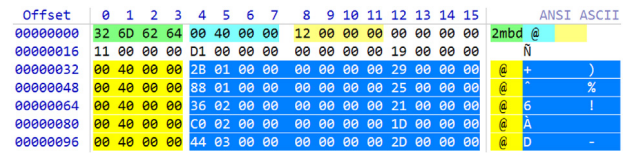

Fig. 2. Example of database map.

**Table 3**
Data page structure.

| Offset | Bytes | Description |
| --- | --- | --- |
| 00 | 4 | Header string recorded as 2mbd |
| 04 | 4 | Page size (physical) |
| 08 | 4 | Size (allocated) |
| 12 | 4 | Record page sub-type |
| 16 | 4 | Size |
| 20 | variable | Data |

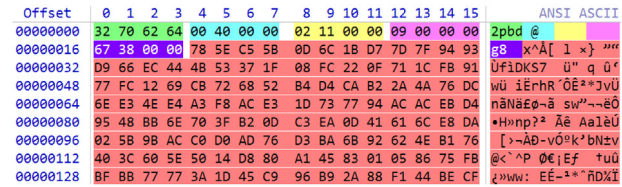

Fig. 3. Example of compressed page.

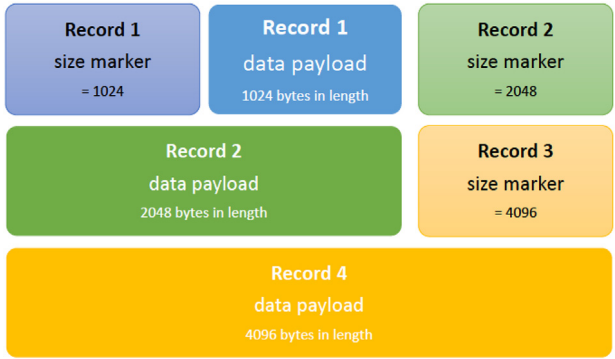

Fig. 4. Example of decompressed page.

Fig. 5. Example of the back-to-back storage of metadata records.

Of particular interest, is the existence of the Catalog Node ID (CNID) and Parent CNID markers found within the records. The metadata store shows similarities in structure to a network database, where each record maintains a relationship with its parent and therefore allow a hierarchical arrangement of records to be built. The CNID is used by the HFS + filesystem to uniquely identify each file/folder on the system. An important feature is that they are not reused until exhausted (i.e. you reach the structural limit and the count restarts). New files created on an HFS + filesystem are given the next consecutive CNID, even if an earlier one has been made available because of a deletion. Other observations outside the scope of this paper have been made, all of which warrant further investigation and should be the subject of any future

research. A brief summary of some of these is presented below:

- The order in which metadata values are listed is different dependant on the operating system used to perform the experiments.
- No clear landmarks have been noted for the metadata records with many of them existing as variable length strings sitting back to back, separated by NULL terminators (00) and/or separated by the hex value 01.
- Empty metadata values are not populated within the record making the size of each record efficient.
- A number of known metadata fields have not been resolved within the record although being returned by queries sent to Spotlight.

*Metadata fields (data page type 17)*

Data found within data page type 17 have been found to contain a record of Metadata Attribute fields. The field records start at offset 32 with the first 4 bytes detailing the record number followed by unknown flags and a variable length string (NULL terminated). We also wrote a python script to analyse this type of page. The output of the script is a CSV, detailing the record number, flags and metadata attribute name.

The order these records appear, show a relationship with the order records are found within the data page. It will be important to understand this relationship in order to be able to parse any recovered records and should be the subject of future research.

Data found within data page type 33 contain a record of Uniform Type Identifier (UTI). The field records start at offset 32 with the first 4 bytes detailing the record number followed by a null terminated string declaring the UTI field. Certain records also contain an additional language/country code field (Fig. 7).

*Persistence of records within store.db*

Experiment 1 and Experiment 2 showed that metadata records within the metadata store persist for a period of time after the corresponding file on the filesystem has been deleted. In addition, large deletion events (where all the files within a folder are deleted), cause database pages to become deleted and therefore recoverable from unused space on the filesystem — this is likely more prevalent with TRIM on SSD devices (Vieyra et al., 2018).

A summary of the results from Experiment 1 is detailed in Fig. 8. The file deletion events within the table have been highlighted in red and file creation events highlighted in green. All events where a deletion and creation occurred between snapshots have been highlighted in yellow.

It was noted that although Spotlight reported that indexing was complete and the MDS process was idle, not every file was indexed. When files are added to the system or deleted, the. store.db is populated first followed by the store. db§ and it may take several minutes to pass before the metadata store is completely updated. This is consistent with FSEvents notifications waiting in a queue for MDS to act upon. After a period of time has passed, however, the databases catch-up with all outstanding notifications.

Fig. 6. Breakdown of a metadata store record for a word document.

Fig. 7. Structure of uniform type identifiers.

| Snapshot | Items added or removed | macOS 10.12 | | OS X 10.11 | | OS X 10.8 | | macOS 10.12 | OS X 10.11 | OS X 10.8 |
|---|---|---|---|---|---|---|---|---|---|---|
| | | store.db records | .store.db records | store.db records | .store.db records | store.db records | .store.db records | unallocated records | | |
| 00 | | 185882 | 185882 | 165116 | 165115 | 127423 | 127519 | 0 | 0 | 0 |
| 01 | +885 | 185988 | 185988 | 165315 | 165906 | 127722 | 127746 | 0 | 0 | 0 |
| 02 | -502 | 185988 | 186936 | 165315 | 165906 | 127722 | 127746 | 0 | 0 | 0 |
| 03 | | 186434 | 186434 | 165560 | 165560 | 127722 | 127746 | 0 | 0 | 0 |
| 04 | | 186434 | 186434 | 165560 | 165560 | 128038 | 128038 | 0 | 0 | 0 |
| 05 | | 186434 | 186434 | 165560 | 165562 | 128038 | 128040 | 0 | 0 | 0 |
| 06 | | 186434 | 186434 | 165568 | 165568 | 128143 | 128145 | 0 | 0 | 0 |
| 07 | | 186434 | 186435 | 165568 | 165568 | 128143 | 128145 | 0 | 0 | 0 |
| 08 | +1215 | 186434 | 186435 | 165568 | 167995 | 128143 | 130600 | 0 | 0 | 0 |
| 09 | -740 | 186434 | 187907 | 165568 | 166691 | 129112 | 129111 | 0 | 0 | 1425 |
| 10 | | 186434 | 187167 | 165568 | 166676 | 129112 | 129112 | 717 | 0 | 1425 |
| 11 | | 186434 | 187167 | 165568 | 166684 | 129112 | 129111 | 345 | 0 | 1425 |
| 12 | | 186434 | 187167 | 166684 | 166684 | 129111 | 129111 | 0 | 0 | 1426 |
| 13 | | 186434 | 187167 | 166684 | 166684 | 129111 | 129111 | 0 | 0 | 1426 |
| 14 | | 187167 | 187167 | 166684 | 166684 | 129111 | 129111 | 0 | 0 | 1426 |
| 15 | +1639 | 187167 | 188780 | 166690 | 166468 | 131145 | 130083 | 1001 | 1902 | 1171 |
| 16 | | 187167 | 188780 | 166690 | 167494 | 131145 | 129924 | 957 | 1902 | 1171 |

**Fig. 8.** Summary of experiment 1 results.

What is evident, is that a number of database pages were deleted from the database during mass deletion events and these records are recoverable from unused space on the filesystem. Keyword searches were constructed using the names of deleted files and run on the decompressed records within the metadata store. The purpose was to identify metadata records for deleted files persisting within the databases. A summary of the findings is provided below:

- The database is populated with metadata records of extant files on the filesystem.
- Once the file is deleted, the record was also deleted within 5 min of the event.
- No records are recoverable from unused space on the database once it has been deleted.
- The records only exist within the allocated space of the database page.

Database pages containing hits were manually spot checked to identify where in the page they existed. In each case, the unused space of the page had been wiped (00) and the records were no longer recoverable.

As shown earlier, each record is of variable size and sits back-to-back with the next record. There are no markers defining the location of records. The data stored within the pages are compressed and when records are deleted, the remaining records show fluidity and collapse into the newly made available space. This operation overwrites any deleted records. Fig. 9 shows this fluidity of records and offers an explanation of why deleted records are not recoverable from within the database.

The results from Experiment 2 are summarised in Fig. 10. The 'count' column is used to display the expected number of records within the store databases and is calculated as: Number of Extant Files + Number of Extant Folders +2.

In general, the experiment behaved as would be expected: The store databases were populated as files were added and de-populated when files were deleted. It was confirmed that once a record is deleted in the store database it is no longer recoverable. The noticeable exception is the store. db located within the FAT32 formatted volume. Every single metadata record remained intact within the store. db and is recoverable. The records remained intact for the duration of the experiment. When all of the files were deleted, it caused the data pages within the store. db database to become unallocated on the filesystem and therefore also recoverable by Forensic Investigators.

*Persistence of records between two operating system*

In Experiment 3, two USB devices (formatted as FAT32 and exFAT respectively) were moved between a macOS system and a Windows 10 system. Each time the drive was reconnected to the macOS system, no other action would be performed for 5 min before a snapshot being created. A summary of the results is provided in Fig. 11.

On review of the results, it was noted that the initial process of adding files to the USB devices and allowing Spotlight to complete indexing did not work. At snapshot 01 there was an expectation that the databases would contain 2753 records, however, only half that amount was actually indexed. This has skewed all remaining steps and made the analysis difficult. What can be confirmed from the experiment is summarised as: (i) Records of deleted files were consistently found within the metadata stores and would persist between connections to the Mac OS X computer; (ii) The deletion of files on a Windows computer resulted in a number of database pages becoming unallocated when the devices were re-connected to the macOS computer.

*Erasing the spotlight metadata store*

The experiments; Experiment 4, Experiment 5 and Experiment 6 showed that deleting or causing the Spotlight index to be refreshed resulted in the metadata store records being deleted and available for recovery from unused space on the filesystem. Summaries of the results are shown in Fig. 12, Fig. 13 and Fig. 14

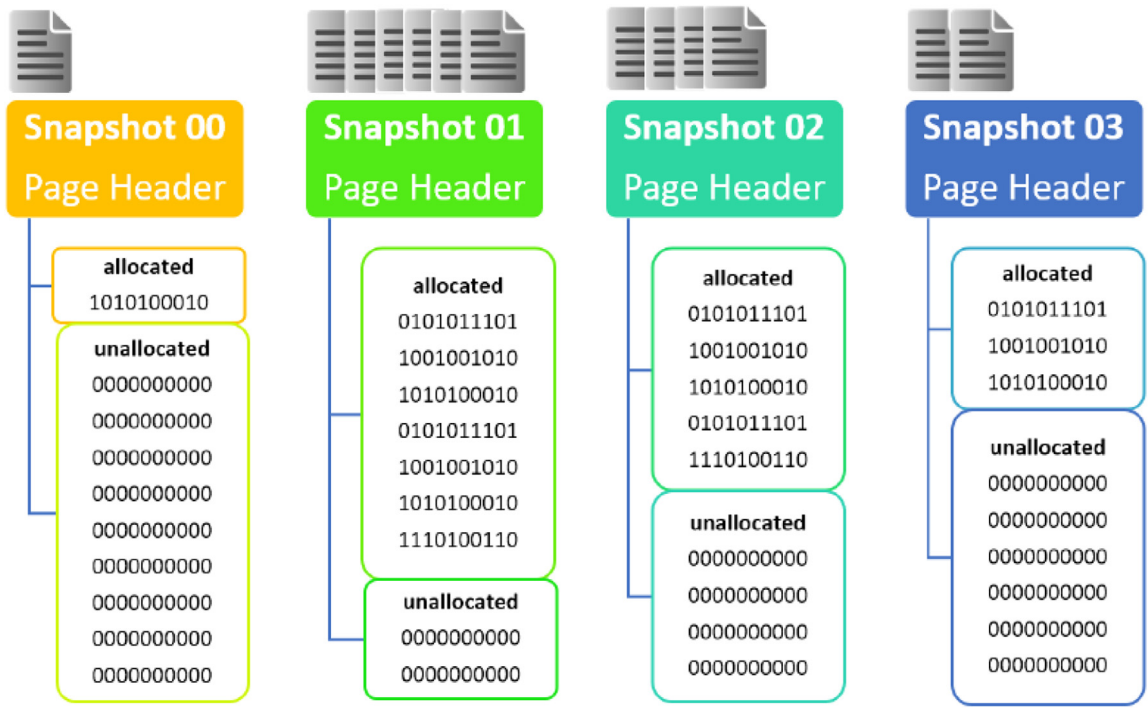

Fig. 9. Page allocation-records are created or deleted.

| Snapshot | Count | exFAT | | FAT32 | | HFS+ | | exFAT | FAT32 | HFS+ |
| | | store.db records | .store.db records | store.db records | .store.db records | store.db records | .store.db records | unallocated records | | |
|---|---|---|---|---|---|---|---|---|---|---|
| 00 | 2 | 2 | 2 | 2 | 2 | 2 | 2 | 0 | 0 | 0 |
| 01 | 2753 | 2738 | 2738 | 2753 | 2753 | 2753 | 2753 | 0 | 0 | 0 |
| 02 | 350 | 350 | 350 | 2753 | 350 | 350 | 350 | 0 | 0 | 0 |
| 03 | 350 | 350 | 350 | 2753 | 350 | 350 | 350 | 0 | 0 | 0 |
| 04 | 350 | 350 | 350 | 2753 | 350 | 350 | 350 | 0 | 0 | 0 |
| 05 | 350 | 350 | 350 | 2753 | 350 | 350 | 350 | 0 | 0 | 0 |
| 06 | 350 | 350 | 350 | 2753 | 350 | 350 | 350 | 0 | 0 | 0 |
| 07 | 0 | 350 | 100 | 2753 | 100 | 350 | 2 | 35 | 41 | 314 |

Fig. 10. Summary of experiment 2 results.

| Snapshot | Count | exFAT | | FAT32 | | exFAT | FAT32 |
| | | store.db records | .store.db records | store.db records | .store.db records | unallocated records | |
|---|---|---|---|---|---|---|---|
| 00 | 2 | 2 | 2 | 2 | 2 | 0 | 0 |
| 01 | 2753 | 7 | 1526 | 7 | 1480 | 0 | 0 |
| 02 | 377 | 483 | 483 | 392 | 392 | 107 | 380 |
| 03 | 377 | 483 | 483 | 392 | 392 | 107 | 380 |
| 04 | 2 | 483 | 481 | 392 | 390 | 107 | 380 |
| 05 | 157 | 125 | 125 | 24 | 24 | 107 | 778 |
| 06 | 157 | 130 | 130 | 24 | 24 | 107 | 778 |

Fig. 11. Summary of experiment 3 results.

respectively.

In each experiment, the process of re-indexing the Spotlight metadata store, either through the GUI interface or issuing the mdutil -E command, resulted in the metadata store pages being deleted and available for recovery from unused space on the filesystem. Each time the databases were deleted, the store databases

| Filesystem | Notes | store.db records | .store.db records | unallocated records | Snapshot |
|---|---|---|---|---|---|
| HFS+ | Standard build | 185882 | 185882 | 0 | 00 |
| HFS+ | 2876 extant files 24 folders | 188922 | 186160 | 0 | 01 |
| HFS+ | 2876 extant files 24 folders | deleted | deleted | 375082 | 02 |

Fig. 12. Summary of experiment 4 results.

| Snapshot | store.db records | .store.db Records | unallocated records |
|---|---|---|---|
| 00 | 185882 | 185882 | 0 |
| 01 | Database not available | Database not available | 371912 |
| 02 | Database not available | Database not available | 378015 |
| 03 | Database not available | Database not available | 416192 |
| 04 | Database not available | Database not available | 392192 |
| 05 | 188950 | 188950 | 108017 |

Fig. 13. Summary of experiment 5 results.

| Snapshot | CNID store.db | CNID .store.db | store.db records | .store.db records | unallocated records |
|---|---|---|---|---|---|
| 00 | 440303 | 440304 | 185882 | 185882 | 0 |
| 01 | 461015 | 461016 | 186231 | 186231 | 66329 |
| 02 | 458655 | 458656 | 186039 | 186039 | 371317 |
| 03 | 462371 | 462372 | 186984 | 186984 | 392753 |
| 04 | 465046 | 465047 | 189011 | 189011 | 406146 |
| 05 | 465200 | 465201 | 189011 | 189011 | 757481 |

Fig. 14. Summary of experiment 6 results.

| Snapshot | 01 – indexing complete | 02 - mdutil -E command run |
|---|---|---|
| Path | /Macintosh HD/.Spotlight-V100/Store-V2/8BFF9113-83E2-4AC3-8A90-0A1C685A988A | /Macintosh HD/.Spotlight-V100/Store-V2/DeadFiles |
| CNID | 440265 | 463851 |
| Directory Structure | .Spotlight-V100 / Store-V1 / Store-V2 / 8BFF9113-83E2-4AC3-8A90-0A1C685A988A / Cache | .Spotlight-V100 / Store-V1 / Store-V2 / DeadFiles / orphan.87eedc01 |
| store.db records | 188922 | Deleted |
| .store.db records | 186160 | deleted |
| unallocated records | 0 | 375082 |

Fig. 15. Evidence of spotlight deletion recorded in experiment 7.

- Using the mdutil command resulted in the creation of a folder named DeadFiles where the Spotlight volume directory previously existed.
- This is a new folder (not the Volume GUID directory renamed) as confirmed by the different CNIDs.
- It is not known if this behaviour is replicated when the GUI interface is used to regenerate the metadata store.

were created afresh (as confirmed by the change in CNIDs). Evidence of the actual deletion process was captured in Experiment 7, is summarised below and within Fig. 15.

*Recovery of records on casework items*

Experiment 8 showed that deleted metadata records are recoverable from unused space on the filesystems of casework items. A total of 14 forensic images of hard disk drives were examined, with all of the items coming from real world cases. The unused space for each filesystem was searched for deleted database pages and once found were processed. A summary of the results observed is shown in Fig. 16.

Every database page recovered was found to be of size 16,384

| URN | OS X Version \ Source[7] | Pages recovered | unallocated records |
|---|---|---|---|
| 01 | Time Machine Backup Drive | 314 | N/A |
| 02 | Time Capsule | 1045 | N/A |
| 03 | Mac OS X 10.4.6 | 0 | |
| 04 | Mac OS X 10.5.8 | 3022 | N/A |
| 05 | Mac OS X 10.5.8 | 21 | N/A |
| 06 | Mac OS X 10.6.8 | 405 | 53,855 |
| 07 | Mac OS X 10.6.8 | 143 | 87 |
| 08 | Mac OS X 10.7.5 | 3477 | 61,808 |
| 09 | Mac OS X 10.7.5 | 852 | 7,518 |
| 10 | Mac OS X 10.8.5 | 526 | 219757 |
| 11 | Mac OS X 10.9.5 | 1406 | 461917 |
| 12 | Mac OS X 10.9.5 | 1303 | 300142 |
| 13 | Mac OS X 10.12.6 | 2019 | 490,130 |
| 14 | Apple hard disk drive formatted for use on a windows system | 1004 | 255,724 |

**Fig. 16.** Summary of experiment 8 results.

bytes and always existed at a sector boundary (a result of the page size being a multiple of 2 and greater than 512 bytes). In all but one case, deleted database pages were successfully recovered. Although the scope of this research has not included the Store-V1 metadata stores, five of the images examined contained this version of the metadata store databases. When encountered, database pages were still recoverable, however, the records themselves were not processed by the python scripts as they have been developed for Store-V2 databases.

Another hard disk drive was also included in this experiment. This drive when seized was being used on a Windows 10 computer as a storage device. It originated from an Apple Macintosh computer (confirmed by the make and model of the drive). The drive itself had been reformatted as NTFS, however, deleted database pages were successfully recovered, containing over 250,000 records.

### Discussion

In the related work section of this paper, an argument to define the basic structure of the Spotlight metadata store was made, in order that experimental results could be resolved and validated. The metadata store database has been analysed and experiments used to reveal the structure of the metadata store. The work presented has revealed the persistence of metadata records for a period of time within the database that once deleted, are no longer recoverable. In addition, it has been confirmed that deleted pages from the database are recoverable from unused space on the filesystem. As a direct result, the structure has allowed the development of scripts that assist the extraction, recovery, and processing of records. There still exists a number of database structures that remain unresolved and future research is required in order to develop fully automated tools that can parse the metadata store directly and records located within unused space. A summary of the findings of this paper are as follows:

- Metadata records persist within the metadata store after files have been deleted on the file system. However, once a record is deleted from the store, it is no longer recoverable. In cases where the entire directory tree of files is deleted, database pages become deleted with the records intact and can be recovered from unused space on the filesystem.
- If the Spotlight index is reset re-indexed or recreated, the whole metadata store is deleted. These database pages can be recovered from unused space on the filesystem.
- If the Operating system undergoes a major update (from one version to another), a copy of the metadata store is created before being deleted. These database pages can be recovered from unused space on the filesystem.

The metadata store on-disk structure has been confirmed as the store. db file. In experiments, it was confirmed that the database is always co-located with a version named. store.db that would be populated first, before the metadata store itself being updated. It is the presence of these two versions that enable the recovery of persistent records of deleted files. Future research should focus on the exact method the two instance's work in relation to each other and other files located within the. Spotlight-v100 folder structure. It is unknown if the two versions are used to circumvent locks placed on the database by the Metadata Server process, or the. store.db structure acts a write ahead log with transactions being updated during quiet periods.

The basic page configuration of the metadata store database has been described and the relationship between the three-page types has been identified (metadata, UTI, data records), however, the method these tables interact with each other has not been empirically proven. Both the metadata and UTI tables appear to use a primary key to identify their records and there is consistency in the order of fields within the records that is used rather than an explicit field declaration. These observations have a direct impact on the current usability of any recovered records from unused space on the filesystem. It is essential that this association between metadata table and records table is established before the data records can be accurately reported.

In Experiment 3, the initial indexing of files resulted in an incomplete picture of what was occurring. Whilst a USB device is connected to a Windows 10 computer, FSEvents would be unaware of any changes at the directory level and therefore the deletion of files would not register within the metadata store. It is of particular interest why the metadata records become deleted once the volumes were reconnected to the macOS system. Several suggested causes are described below, however, these are new hypotheses and therefore require testing before they can be proved: (i) The entire mounted volume is re-indexed when reconnected; (ii) The volume is checked for any changes; (iii) There are outstanding FSEvents notifications that cause a re-index to reoccur because the files no longer exist on the devices.

### Conclusion and future work

This research offers a new repository of data that can be used to identify the historic values of file attributes and metadata. As files are created and modified on a live system, significant records that provide evidence of a specific event may be overwritten as a result of recent activity. It is established that each time the operating system undergoes a major version update, records within the metadata store become available for recovery within the unallocated database pages residing within the unused space on the filesystem. Investigators should verify that the records are available for recovery and a simple scan of the system logs will confirm if the operating system has been updated. Historic records can then be carved from the filesystem unused space, processed and identified by using the unique CNID identifier. This will offer investigators an opportunity to compare the contents of the recovered record with the current record and would reveal any changes to metadata that occurred between the intervening periods. Examples of the changes include; the movement of the file from one location to another (change in the parent CNID record). The file being renamed, historic timestamps, metadata records that were removed, etc. Future research should be used to provide verification of the exact date these deleted records were created on.

This research has validated that files deleted on the filesystem may still be recorded within the metadata store as extant records. It is proposed that file carving of the unused space for files based on these records take place in order to associate records within the store to data within unused space. Data carving should recover the file as long as it has not been overwritten, and a statistical comparison of the embedded contents of the recovered file can be compared to the existing metadata record. This analysis would provide filesystem attributes that are lost as a result of the deletion, e.g. file name, path and usage timestamps. This type of recovery is novel to the macOS operating system, where historically file attributes are no longer available from the Catalog records once a file is deleted but warning is given that the assignment of attributes possesses considerable risks and the reliability of using such methods should be the topic of future research. There remain several unanswered questions relating to database pages discovered and it is recommended that these form part of any future research: (i) Does the order of the metadata records change? (ii) What happens when new metadata fields are added? (iii) What is the purpose of the unknown data page types?; (iv) What happens if an application is installed that contains new defined metadata fields?; (v) How transferable is the desktop Spotlight investigation technique presented to Apple's mobile operating system, iOS, and to its smart home environment, HomeKit (Goudbeek et al., 2018); (vi) How well can this technique be integrated into centralised cloud forensic processing/Digital Forensics as a Service (DFaaS) 1; (vii) Are there records held within the Content Index Store that are of evidential value?